# Observation of the epsilon-near-zero (ENZ) effect inside the laser plasma in air

Jiayu Zhao

Terahertz Technology Innovation Research Institute, Terahertz Spectrum and Imaging Technology Cooperative Innovation Center, Shanghai Key Lab of Modern Optical System, University of Shanghai for Science and Technology, Shanghai 200093, China

*zhaojiayu@usst.edu.cn

**Abstract**

In the world of epsilon-near-zero (ENZ) materials, the plasma is unique for its natural ENZ properties at the plasma frequency ($\omega_p$). However, for the air plasma during femtosecond laser filamentation with $\omega_p$ in terahertz (THz) band, which is also known as a broadband THz emitter, the possible ENZ effect has long been neglected. In this work, interactions between the laser plasma in air and the radiated THz waves were investigated, and the THz resonance absorption at the ENZ point, which gave rise to the generation of surface plasmon waves, has been theoretically and experimentally demonstrated for the first time, to the best of our knowledge. Specifically, this ENZ effect was accompanied with THz modulations along the plasma filament in multiple domains, including the linear-to-elliptical THz polarization conversion and the multi-ring pattern of far-field THz profiles, etc. Thanks to these novel ENZ-induced phenomena, the understanding of this important plasma-based THz source has been enriched. Moreover, by extending the laser plasma to the role of ENZ material, it is promising to expedite related applications benefiting from the ENZ nature, e.g., the strong spatial confinement of THz waves inside the ENZ region between the bi-filaments array, which realized all-optical time-domain integration of broadband THz pulses as we recently displayed in [arXiv].

## 1. Introduction

Epsilon-near-zero (ENZ) materials [1-5], whose real part of the dielectric constant ($\varepsilon_r$) is approximately zero, represent splendid performances ranging from electromagnetic energy squeezing and tunneling [6,7], slow light trapping [8], electric fields and optical nonlinearity enhancement [9-12], laser harmonics generations [13-15], and the efficient frequency conversion [16,17], etc., hence attracting intense attention in recent years for creating new ENZ materials and applications. One widely used method to realize the ENZ mechanism is to adjust the free carrier concentration of semiconductors [18], metals [5] or transparent conducting oxides [4] for making the collective motion of free carriers at the plasma frequency $\omega_p$.

By contrast, as for a plasma filament induced by femtosecond laser ionization in air without solid substrate materials, it is naturally an ENZ material accessible with no need of plasma density (or free electron density) $N_e$ adjustments for the following two reasons. Firstly, its $N_e$ of around $10^{16\text{-}17}$ cm$^{-3}$ coincides with the cut-off density $N_c$ [19], which enables the ENZ effect at the plasma frequency $\omega_p$ in terahertz (THz) band. Secondly, the plasma filament itself generates and propagates broadband THz waves. Thus an air plasma filament could in principle be an

ENZ-plasmonic structure at THz frequencies for supporting, e.g., surface plasmon waves (SPWs) without external injections of THz pulses.

However, this potential ENZ effect has long been neglected. On one hand, as for the constantly used scheme of two/multi-color laser pumping, although intense THz waves are generated from the plasma filament due to the additional introductions of the laser harmonics, the resultant THz spectroscopy is complicated. For example, the relative phase difference [20,21], relative polarization [22,23] and frequency ratio [24], etc., between the fundamental light and harmonics seriously reshape the THz spectrum along the filament, which might blur the ENZ-related phenomena. On the other hand in single-color case, the emitted THz pulses are generally detected in setup for longitudinal waves, whose far-field profiles have been well interpreted by the quasi-Cherenkov model [25]. And the possible ENZ effects (e.g., THz-SPWs) have not been fully considered due to the absence of transverse THz waves detection.

Fortunately, it has been recently confirmed that transverse THz waves can also be generated by the single-color filament [26,27], which holds promise for revealing the proposed ENZ effect within the laser plasma. Hence in this work, we focused on the single-color driven approach, recorded the transverse THz waves, and observed ENZ-induced phenomena like the linear-to-elliptical THz polarization conversion and the multi-ring pattern of far-field THz profiles. By exploring the laser plasma filament from the ENZ point of view, this Letter enriches our understanding of THz wave generation and propagation during laser filamentation. Furthermore, with the air plasma brought into the scope of ENZ materials, it is promising to expedite associated applications benefiting from the ENZ nature [28].

Compared with previous reports, such as the spatial THz confinement inside the plasma filament [29], which can be simply attributed to the sub-diffraction guiding principle borrowed from the nanophotonic theory [30,31] due to the large THz wavelength (sub-millimeter) versus the small filament diameter (~50 μm), the current work has actually discovered the underlying mechanism for this phenomenon as the fundamental ENZ effect. It is also noticed that the laser-driven gas plasma has been treated as a THz plasmonic particle [32], by which means the THz emission spectrum can be significantly broadened. While in our case, the generated long filament was studied as a THz line-emitter, and we concentrated on the $\omega_p$ spectral absorption due to the ENZ effect rather than the spectral width variations.

## 2. Theory of the ENZ effect inside the laser plasma in air

As shown in Fig. 1(a), in the plasma region induced by femtosecond laser ionization in air, theoretically, the transmission media for THz waves include the ambient air, the plasma filament and the interface between them (i.e., the periphery of plasma column). Corresponding THz dispersion curves ($\omega{\sim}k$) in the three media can be calculated by (i) $k_{\mathrm{THz}}^{\mathrm{air}} = n_{\mathrm{THz}}^{\mathrm{air}} \frac{\omega}{c}$; (ii) $k_{\mathrm{THz}}^{\mathrm{plasma}} = n_{\mathrm{THz}}^{\mathrm{plasma}} \frac{\omega}{c}$, where $n_{\mathrm{THz}}^{\mathrm{plasma}} = \sqrt{\varepsilon_{\mathrm{THz}}^{\mathrm{plasma}}} = \sqrt{1 - \frac{\omega_p^2}{\omega^2 - i\gamma\omega}}$, $\omega_p = \sqrt{\frac{e^2 N_e}{m_e \varepsilon_0}}$, $N_e$ denotes the plasma density, and $\gamma$ corresponds to the typical electron collision frequency inside filament; and (iii) $k_{\mathrm{THz}}^{\mathrm{SPW}} =$

$\frac{\omega}{c}\sqrt{\frac{\varepsilon_{THz}^{plasma}}{\varepsilon_{THz}^{plasma}+1}}$. By further setting $n_{THz}^{air}$ = 1.00027 [33], $N_e$ = 5×10$^{16}$ cm$^{-3}$ and $\gamma \sim$ 1 THz [34], the calculated dispersion curves are drawn in Fig. 1(b) as the black line ($k_{THz}^{air}$), the red line (Re[$k_{THz}^{plasma}$]) and the blue line (Re[$k_{THz}^{SPW}$]), respectively. Three intersection points (yellow circles) along the blue line indicate potential phase matching conditions for the creation of THz-SPWs.

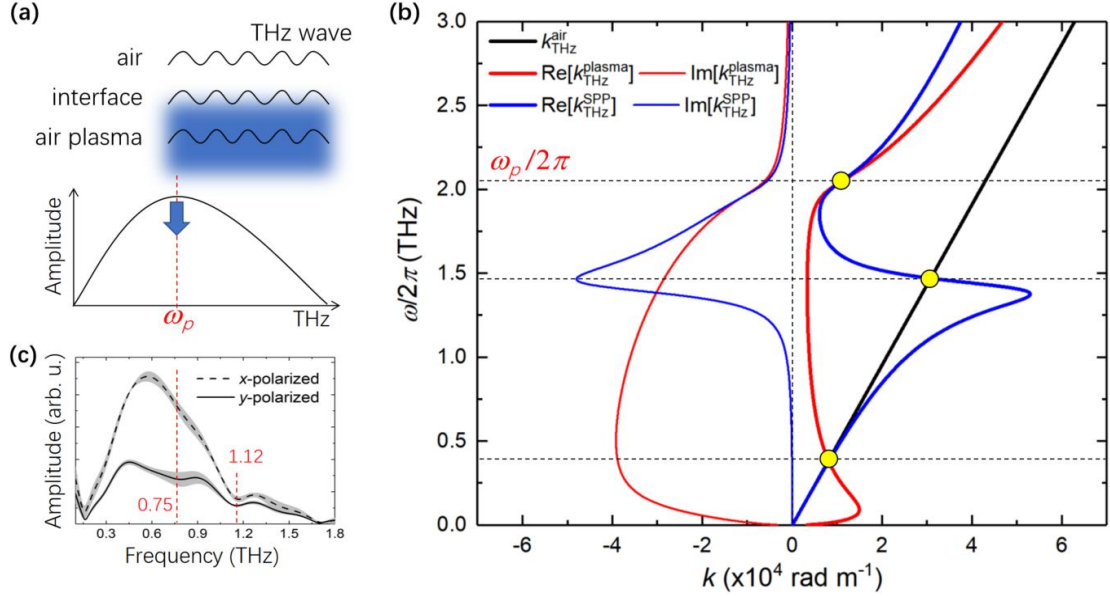

Fig. 1 (a) Propagation of THz waves within three areas of the laser plasma in air, and the typical THz radiation spectrum from the filament which is centered at the plasma frequency $\omega_p$. (b) Dispersion curves of THz waves inside the air, the plasma and the air-plasma interface, respectively. (c) The experimentally detected THz spectra in orthogonal polarization directions from the whole plasma filament with $\omega_p \sim$ 0.75 THz.

On the other hand, modal losses in form of Im[$k_{THz}^{plasma}$] (red thin line) and Im[$k_{THz}^{SPW}$] (blue thin line) need also to be small at the crossing points, avoiding THz-SPWs being immediately absorbed [35]. Taking this issue into account, only the yellow point at the largest $\omega$ of the three could lead to the THz-SPWs ($k_{THz}^{SPW}$) generation with the energy transferred from THz waves inside the plasma ($k_{THz}^{plasma}$). Coincidentally, this phase matching ($Re[k_{THz}^{SPW}] = Re[k_{THz}^{plasma}]$) can only be resulted from $Re[\varepsilon_{THz}^{plasma}] = 0$ at the plasma frequency $\omega_p$, which is also the ENZ point for an air plasma. Thus we used the name of ENZ-SPWs in the following text, since here the ENZ and SPW conditions are simultaneously fulfilled.

Remind ourselves that $\omega_p$ is also approximately the peak frequency for THz emission from the laser filament in air [36], as schematically shown in the lower part of Fig. 1(a). Therefore, when THz waves propagate inside the plasma filament, the THz energy at the vicinity of $\omega_p$ is expected to be converted into the non-radiating ENZ-SPWs, thus likely leaving a valley on peak of the output THz spectrum. This hypothesis has been roughly confirmed as follows. We recorded the transverse THz signal (x- and y-polarized spectral amplitude) radiated from the laser filament, which was created by focusing a femtosecond laser beam (800 nm, 50 fs, 1 kHz and 1.5 mJ/pulse) with a lens (f = 100 cm) in air. It can be seen from the experimental results in Fig. 1(c) that a spectral dip appears at the vicinity of 0.75 THz (more clearly on the *y*-polarized signal), which coincides with the reported $\omega_p$ value [36] near the spectral peak as expected. By contrast, the spectral valley at around 1.12 THz is attributed to the THz energy loss caused by the water vapor absorption in air (Supporting Materials). Apart from the above $\omega_p$ dip, more evidences are provided in the following experiments for further supporting the THz-ENZ-SPW theory.

## 3. Temporal shift and polarization conversions of THz waves

In order to investigate detailed processes of ENZ-SPWs generation, we detected the THz signals emitted from different longitudinal positions of the plasma filament via the cut-back (C-B) method [37]. In brief, a thin ceramic plate (0.3 mm in thickness) was normally inserted into the plasma column to terminate the filament, and then moved longitudinally. THz waveforms emitted from this length-varied filament were detected in both *x* and *y* orthogonal directions as shown in Fig. 2(a) and (b), respectively. Here *z* corresponded to the inserting position of the plate, and *z* = 0 mm was approximately the starting point of the filament.

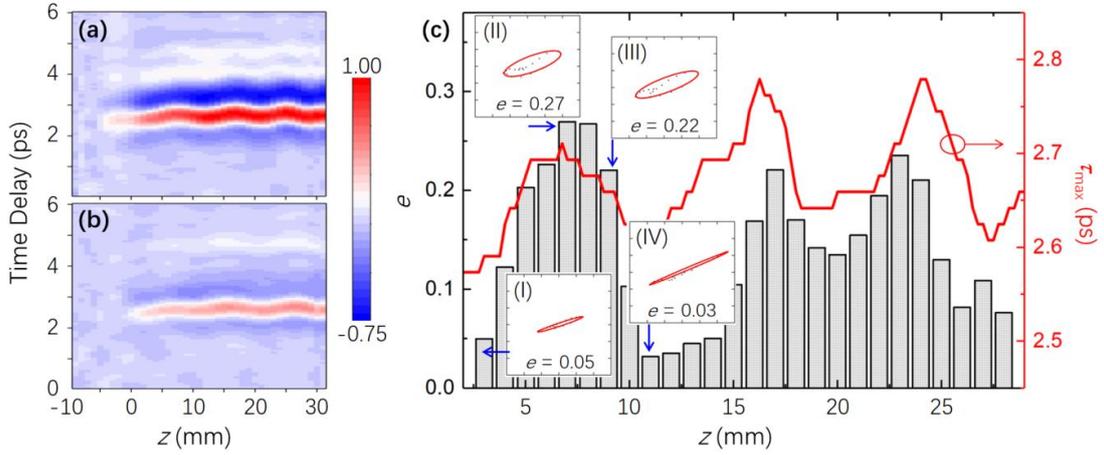

Fig. 2 Temporal THz waveforms emitted from different lengths of the laser filament in (a) *x* and (b) *y* directions, respectively. (c) The evolution of THz polarization ellipticity (bars) along the plasma filament, along with the temporal locations of the THz amplitude maximum (the red line). Insets: (I-IV) Conversions between the linear and elliptical THz polarization along *z*.

One can clearly see that along the *z* axis, there are about three periods of temporal delay and advance for the THz transients. This phenomena of THz waveform displacement in time domain is typically resulted from interactions between THz waves and the laser plasma [26,29,38]. In short, spatial confinements of THz energy inside the plasma filament occurred, and the THz propagation medium switched between air and plasma, altering the effective refractive index of THz frequency components around the unity. Hence the detected THz pulse shifted temporally. This THz-plasma interaction can also be observed from the THz polarization evolution along the filament as shown in Fig. 2(c), inside which the ellipticity ($e = E_{\text{THz}}^{\text{minor axis}} / E_{\text{THz}}^{\text{major axis}}$) of the THz polarization is calculated (as bars) by the THz waveforms in Fig. 2(a) and (b). It can be seen that the *e* values also varied in three periods, along with the red line indicating the temporal location $\tau_{\max}$ of the maximum THz amplitude. In each period, the THz polarization changed between linear and elliptical patterns (inset as I-IV), which was due to the birefringence effect during THz confinement inside the filament [26, 39].

## 4. Modulations of THz spectral amplitude and refractive index at the ENZ frequency

The above THz-plasma interaction is crucially important since it produced $k_{THz}^{plasma}$ for phase matching with $k_{THz}^{SPW}$ as mentioned in Fig. 1(b), and in this way the $\omega_p$-dip-featured THz spectrum can be further realized as shown in Fig. 1(c). Now this issue has been checked during the cut-back (C-B) measurements as follows.

As shown in Fig. 3(a) and (b), the C-B resultant spectra are achieved by Fourier transforms (FT) on the data in Fig. 2 (a) and (b) for the $x$- and $y$-polarization, respectively. Note that, Fig. 3(a) and (b) are normalized to their own maxima for better image contrasts. On the left of both figures, the overall spectral profiles are also plotted as white lines. Furthermore, black and white dashed lines marked the frequency locations of the $\omega_p$ dip [~0.75 THz as (I) and (II)] and the water absorption dip [~1.12 THz as (III) and (IV)].

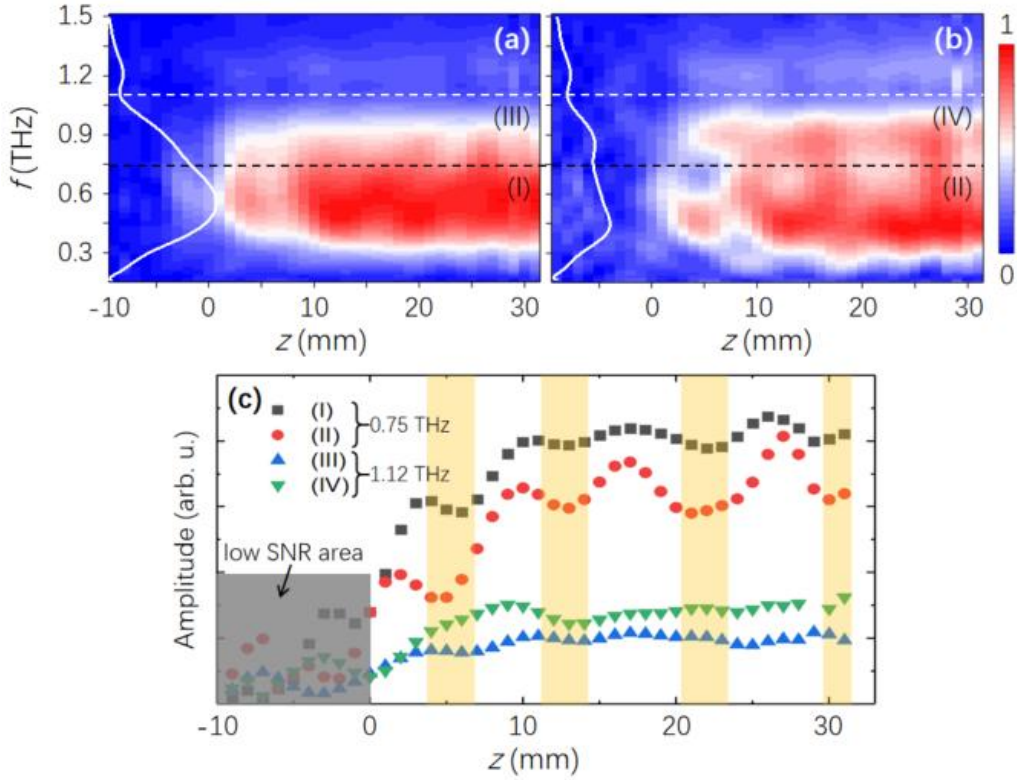

Fig. 3 THz spectral amplitude emitted from different lengths of the laser filament in (a) $x$ and (b) $y$ directions, respectively. (c) The variation of amplitude at 0.75 and 1.12 THz along the dashed lines (I-IV) in (a) and (b).

The spectral amplitude along (I)-(IV) are extracted and drawn in Fig. 3(c). It can be clearly seen that, (I) and (II) at around 0.75 THz have several valleys along $z$ axis (in yellow shadow areas), which match the $z$-periods of temporal shifts of the THz waveforms in Fig. 2. More importantly, the valleys correspond to the reductions of THz energy, which was transferred to ENZ-SPWs as presented in Fig. 1(c). By contrast with (I-II), (III-IV) at about 1.12 THz remained almost the same during C-B operations, which is due to the fact that they were induced by the constant water absorption without periodicity.

Next, the refractive indices of THz emission (especially at the ENZ frequency, i.e., $\omega_p$) are investigated for further proving the creation of ENZ-SPWs, since when resonance energy transfer occurs at a certain frequency, the local index will undergo shift. In order to retrieve this index change, firstly, we obtained the THz phase distributions as shown in Fig. 4(a) via FT on Fig. 2(a). Along the red dashed line at $\omega_p/2\pi \sim 0.75$ THz, a representative phase evolution ($\varphi$) as a function of $z$ is extracted and plotted in the upper part of Fig. 4 as red open circles. Obviously, three periodic fluctuations of $\varphi$-$z$, which correspond to the temporal delay (TD) or temporal advance (TA) of THz pulses, are in agreement with that of Fig. 2.

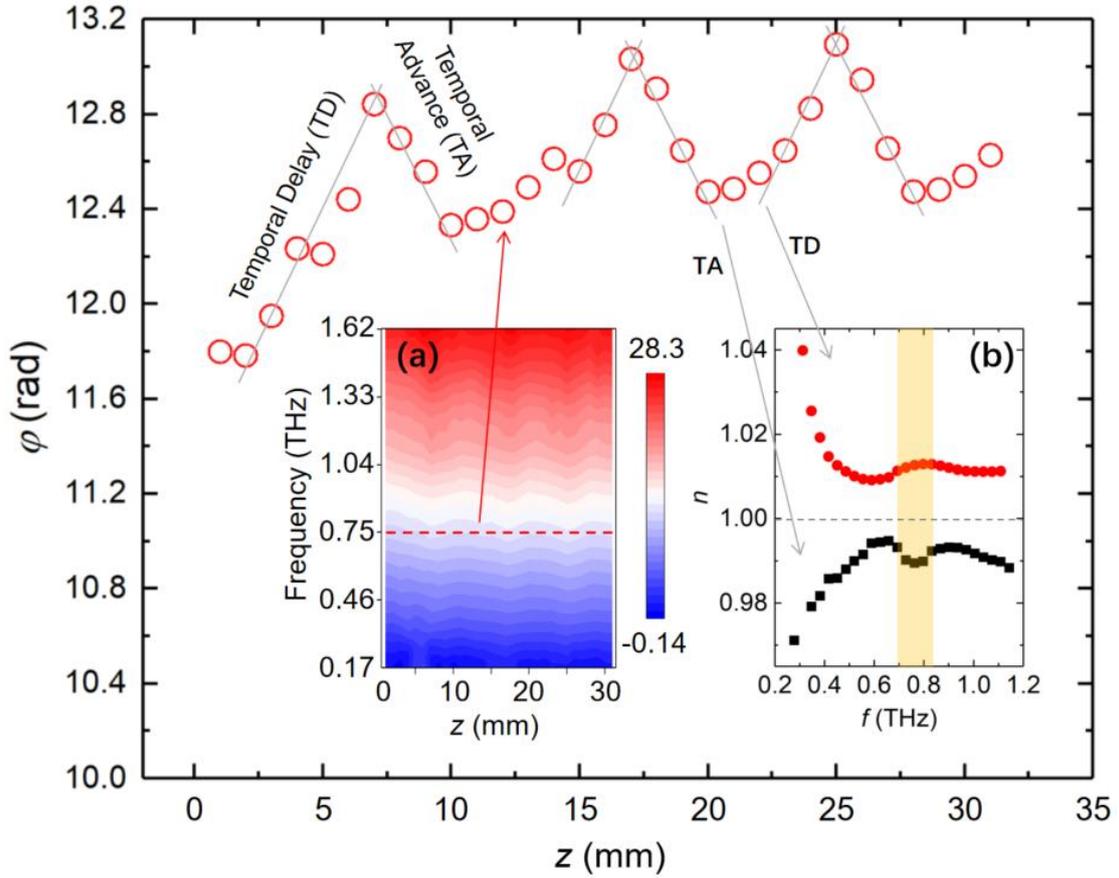

Fig. 4 (a) Distributions of THz spectral phase along the filament. The phase variations at 0.75 THz (along the red dashed line) are plotted as red open circles. (b) The calculated THz refractive indices given by the phase variations in (a).

Secondly, the refractive index at $f = 0.75$ THz can be calculated by this $\varphi$-$z$ relationship with $n_{THz} = (c/\omega)d\varphi/dz + n_{THz}^{plasma}$ [40], where $\omega = 2\pi f$ is the angular frequency, and $d\varphi/dz$ is the slope of the linear fitting (gray lines) in Fig. 4. For the rising gray lines (marked with "TD"), one can resolve a $n_{0.75\ THz}$ slightly larger than 1, and for the falling ones (marked with "TA"), $n_{0.75\ THz}$ is a little smaller than 1. In this way, finally, two phase indices (>1 and <1) for each THz frequency are obtained, and the complete dispersion curves ($n$-$f$) are shown in Fig. 4(b). The observed index jump (between red and black squares around the unity) agrees with the temporal THz pulse shifts (delay or advance) presented in Fig. 2. More interesting, anomalous *n* values can be clearly seen around 0.75 THz (in the yellow shadow area), at which frequency the ENZ-SPWs were generated. Consequently, the ENZ effect has been re-confirmed inside the plasma filament in air.

One may additionally observe that the anomalous *n* mainly exist in the *n* < 1 region in Fig. 4(b), which hints that the conversion to ENZ-SPWs is accompanied by the temporal advance of the THz waveforms (i.e., THz spatial confinement inside the plasma). This result coincides with the ENZ precondition mentioned before, that the THz-plasma interactions lead to $k_{THz}^{plasma}$-$k_{THz}^{SPP}$ phase matching, and accordingly ENZ-SPWs are created. In addition, the calculated $n_{0.75\text{ THz}}$ (in the <1 region) changed towards smaller values as shown in Fig. 4(b), which fits the fact that the refractive index at the ENZ point is little (theoretically down to zero).

## 5. The multi-ring pattern of conical THz radiations in the far field

The above analyses are mainly focused on the near-field properties of THz wave propagation during laser filamentation. In this section, we turned to the far-field case, and the transverse profile of the emitted THz beam has been characterized in the forward direction of the filament. Here, an aperture was positioned after the filament at *z* ~ 60 mm as displayed in Fig. 5(a). Varying the opening radius (*r*) of the aperture, the THz power $P_{THz}(r)$ was recorded as shown as black squares in Fig. 5. It can be clearly seen that the collected THz power experienced growths mainly in three yellow shadow regimes marked with upwards blue arrows. Beyond these radius locations, the THz signal increased relatively slowly. Moreover, $P_{THz}(r)$ is given by the integral equation of $P_{THz}(r) = \int I_{THz}(r) \cdot 2\pi r \cdot dr$, where $I_{THz}(r)$ is the THz intensity at the radial position *r*. Thus, $I_{THz}(r)$ can be retrieved and the results are shown in Fig. 5 as gray bars, whose maximums are also overlaid with the yellow shadow regimes. This $I_{THz}(r)$ distribution actually indicates that the THz yield is mostly within three conical rings.

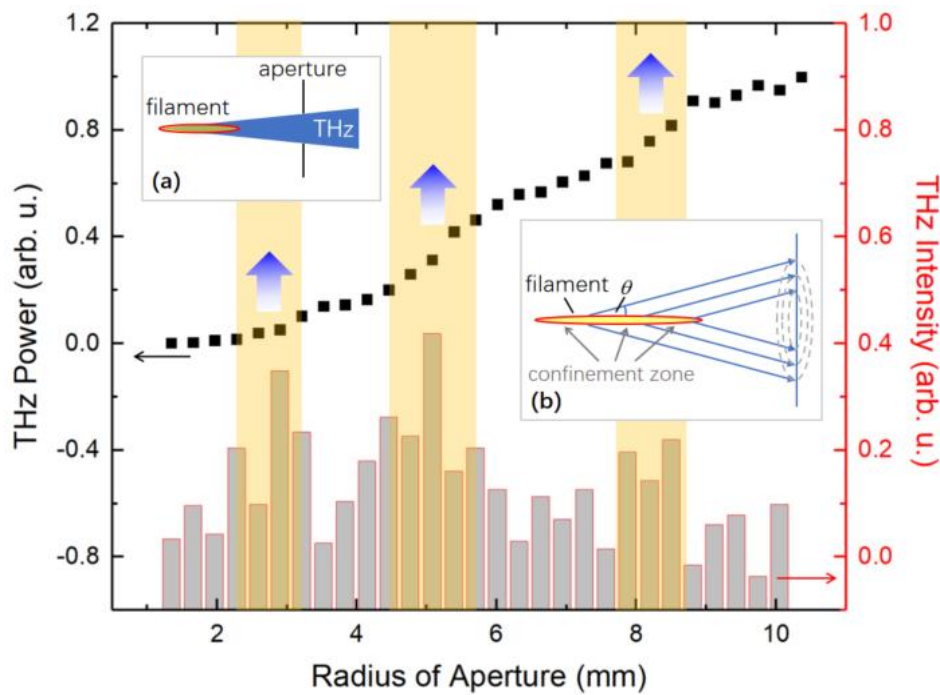

Fig. 5 Far-field THz signals (black squares) and the corresponding differential results (gray bars) as a function of the aperture opening radius, which is schematically shown in (a). (b) Multi-ring conical THz radiations from three longitudinal positions of the laser filament.

The above phenomena are further clarified in Fig. 5(b). Fig. 5(b) shows at first the three times of THz wave confinement during propagation along the filament as mentioned in Fig. 2. When this happened, the THz wave was constrained inside the plasma and its energy was transferred to the nonradiative SPWs, without emitting to the far field. By contrast, beyond the confinement zones, THz waves radiated in the forward direction with similar angular distributions (set as $\theta$) according to Ref. [41]. Then, a connection can be established between the far-field THz radial distribution and the longitudinal emitting positions along the filament, which are separated by THz confinement zones as shown in Fig. 5(b). Hence, in the far field, one can only detect considerable THz intensity at several certain radius positions (i.e., multi-ring, instead of single-ring or bell-shaped distribution), as shown as gray bars in Fig. 5.

## 6. Conclusion

In summary, by studying the transverse THz waves driven by the single-color femtosecond laser field in air, the creation of THz-SPWs at the ENZ point has been uncovered, together with the THz polarization evolution and far-field profile modulation, etc., along the plasma filament. Owing to this THz-ENZ platform, the entire processes of THz wave generation and confined propagation as well as SPWs' conversion are connected. Thus the suggested ENZ effect paves the way towards deep understanding of this important plasma-based THz source. Furthermore, future applications could also benefit from this ENZ-type air plasma, and new discoveries might soon come into being [28].

# Supplementary Materials

## 1. The water absorption of THz waves

As shown in Fig. 1(c) of the main text, besides the $\omega_p$ dip at 0.75 THz, there is another 1.12-THz dip attributed to the absorption of water vapor in air. In order to clarify this point, here we detected the THz spectra with a commercial THz-TDS system (THZ040, EKSPLA). Accordingly, the THz spectra with low humidity of 3.3% (black solid line) and routine humidity of 25% (red dashed line) are shown in Fig. S1. One can clearly see major THz losses in two yellow shadowed areas, which centered at around 1.1 THz and 1.7 THz, respectively. Both frequency locations coincide with the THz amplitude valleys in Fig. 1(c), thus proving the water absorption issue. It is worth noting that, the spectral resolution of our experimental setup used in the main text is not as high as the commercial TDS instrument, so the fine structures of the water absorption lines in Fig. S1 cannot be well resolved in Fig. 1(c) of the main text.

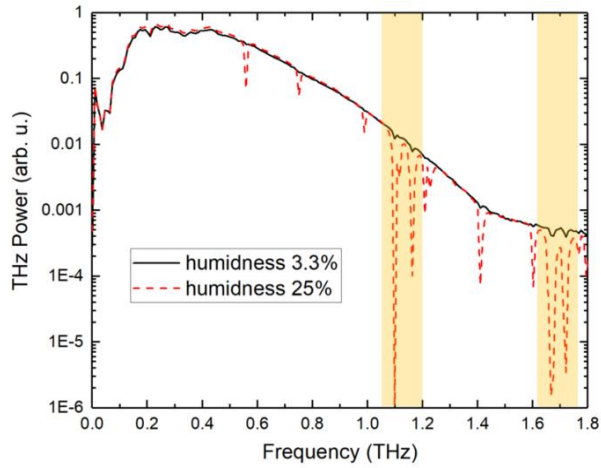

Fig. S1 THz spectra with water absorption at two humidities of 3.3% (black line) and 25% (red dashed line), respectively. The shadowed areas indicate two main frequency locations of THz energy loss, which agree with Fig. 1(c) of the main text.

## 2. Non-radially polarized THz components from the laser filament

It is worth noting that, as shown in Fig. 2(a-b) of the main text, the $x$-polarized THz amplitude obviously exceeds the $y$-polarized one emitted from the single-color laser filament, thus leading to the polarization state (Fig. 2(c) I-IV) with a near-$x$-axis azimuth angle (~20 degree) and a small ellipticity ($e < 0.27$). This result agrees with previous reports [S1, S2], and can be interpreted by the following two points.

Firstly, the laser ponderomotive force in the transverse direction drove the plasma electron away from the $z$ axis, and the resulted radial current [S3] (rather than the longitudinal current) gave birth to this transverse THz radiation from the single-color filament. Secondly, according to Ref. [S4, S5], the linear $x$-polarization of the driven laser used in our experiments affected differently along $x$ and $y$ axes on (i) the air ionization, and (ii) the following electron motions, as well as (iii) the THz generation. Both the above issues gave rise to this transversely linear/elliptical THz radiation as detected in this work.